\documentclass[aps,12pt,preprintnumbers,nofootinbib,superscriptaddress]{revtex4}

\usepackage{graphicx}
\usepackage{amssymb,amsmath,slashed}

\newcommand\Tr{\text{Tr}}

\newcommand\beq{\begin{equation}}
\newcommand\eeq{\end{equation}}

\begin{document}


\preprint{UCSD/PTH 07-11}

\title{One-Loop Renormalization of Lee-Wick Gauge Theory}
\author{Benjam\'in Grinstein}
\email[]{bgrinstein@ucsd.edu}
\affiliation{Department of Physics, University of California at San Diego, La Jolla, CA 92093}

\author{Donal O'Connell}
\email[]{donal@ias.edu}
\affiliation{Institute for Advanced Study, School of Natural Sciences, Einstein Drive, Princeton, NJ 08540}

\date{\today}

\begin{abstract}
We examine the renormalization of Lee-Wick gauge theory to one loop
order. We show that only knowledge of the wavefunction renormalization
is necessary to determine the running couplings, anomalous dimensions,
and vector boson masses. In particular, the logarithmic running of
the Lee-Wick vector boson mass is exactly related to the running
of the coupling. In the case of an asymptotically free theory,
the vector boson mass runs to infinity in the ultraviolet. Thus, the 
UV fixed point of the pure gauge theory is an ordinary quantum field
theory. We find that the coupling runs more quickly in Lee-Wick gauge
theory than in ordinary gauge theory, so the Lee-Wick standard model
does not naturally unify at any scale. Finally, we present results on
the beta function of more general theories containing dimension six
operators which differ from previous results in the literature.

\end{abstract}

\maketitle

\section{Introduction}
In recent months, an extension of the standard model of particle
physics has been constructed~\cite{us} based on ideas of Lee and
Wick~\cite{leewick}. Lee and Wick constructed a finite theory of
quantum electrodynamics in order to remove divergences in certain
mass corrections. The theory of Lee and Wick contains new degrees
of freedom which are associated with wrong sign kinetic terms. Thus
the theory is classically unstable. Lee and Wick proposed that the
instability could be removed at the classical level by imposing
boundary conditions on the theory, and at the quantum level
by quantizing the theory such that the energy of any scattering
(asymptotic) state is positive. This requires the introduction of a non-positive
definite norm on the Hilbert space. Lee and Wick further described how
the theory could nevertheless be unitary if the negative norm states
are heavy and can decay to states
of positive norm. These ideas have been discussed extensively in the
literature~\cite{leewick,Coleman,CLOP,Nakanishi:1971jj,Lee:1971ix,Nakanishi:1971ky,Antoniadis:1986tu,Boulware:1983vw,Kuti,van Tonder:2006ye}.
It has not been shown that an arbitrary Lee-Wick theory is unitary to
all orders of perturbation theory, but there is no known example of a
theory that cannot be unitarized in this way. In particular, scalar
Lee-Wick theories have been extensively analyzed in~\cite{Kuti} at the
non-perturbative level with encouraging results.

With the modern understanding of renormalization the original motivation
of Lee and Wick is no longer pressing, and in particular the massive
resonances predicted by the Lee-Wick theory of electrodynamics have not
been observed. Thus, interest in the Lee-Wick model of electrodynamics
has dwindled. However, we are currently faced with quadratically
divergent radiative corrections to the Higgs mass. The extension of the
standard model developed in~\cite{us}, known as the Lee-Wick standard
model, includes new degrees of freedom that remove these quadratic
divergences. The resulting theory is logarithmically divergent. The new
degrees of freedom are associated with higher derivative, dimension six
operators present in the microscopic Lagrangian of the theory. It was
shown that an equivalent formulation of the theory contains only dimension
four operators; in this form, the new degrees of freedom in the theory
have wrong sign kinetic Lagrangians. The Lee-Wick prescription is then
invoked to quantize the theory. Physically, the Lee-Wick standard model is
unusual since the future boundary condition leads to acausality. However,
the time scale of this acausality is far too small to have been ruled
out by experiment.

The flavor structure of the Lee-Wick standard model has been explored
in~\cite{Tim} with the attractive result that while new flavor changing
neutral and charged currents are present, the flavor symmetry violation
is naturally within experimental bounds. However, the Lee-Wick standard
model was defined by choosing particular dimension six operators to
add to the standard model Lagrangian. One could imagine a more general
theory containing a greater number of dimension six operators. Some
of these operators would lead to unacceptably large flavor changing
currents. In~\cite{Grinstein:2007iz} the question of the physical status
of such operators was addressed, and it was shown that the choice of
operators made in defining the Lee-Wick standard model is such that
scattering amplitudes in the theory do not violate the well-known
perturbative unitarity bounds. Thus, while dimension six operators
typically imply either strong coupling in the ultraviolet or a violation
of unitarity, the operators included in the Lee-Wick standard model lead
to a perturbative UV completion as suggested by precision
electroweak constraints. Aspects of the LHC phenomenology of the Lee-Wick
standard model has been discussed in~\cite{Rizzo:2007ae,Krauss:2007bz,Rizzo:2007nf},
and more theoretical aspects of these models have been examined
in~\cite{Wu:2007yd} and in~\cite{Espinosa:2007ny}. Supersymmetric models
including similar higher dimension operators have been examined in~\cite{Antoniadis:2007xc}.

In the present work, we turn to the question of the one-loop structure of
non-abelian Lee-Wick gauge theory. A perturbative power counting argument
presented in~\cite{us} establishes that the dimension six operators
in the higher derivative formulation of the theory only receive finite
renormalizations. In this work, we examine the renormalization in more
detail. We work in background field gauge. There are some subtleties
of gauge fixing in these theories which we discuss before turning our
attention to the beta function and anomalous dimensions of matter. One
interesting result is that the running of the massive vector boson mass, $m$,
in the theory is exactly related to the running of the coupling, $g$, because
the quantity $m g$ is a renormalization group invariant. In an
asymptotically free quantum field theory, $g$ runs to zero in the
ultraviolet so if $m g \neq 0$, then the mass $m$ must run to infinity
in the UV. Consequently the UV fixed point of the renormalization group
flow is an ordinary free quantum field theory. We find
that the Lee-Wick standard model does not appear to unify at any energy
scale. The Lee-Wick particles in the theory in fact cause the running
of the coupling to be quicker, so that any putative unification
scale would be rather low. If the unification group were to be
semisimple, this would lead to unacceptably large proton decay, but
this problem can be alleviated~\cite{Willenbrock:2003ca}. We then
turn to more general theories containing dimension six operators
which are not of Lee-Wick type. While these theories do not satisfy
the perturbative unitarity bounds, they have nevertheless been
discussed in the literature as a toy model of higher derivative
gravity~\cite{Fradkin:1981hx,Fradkin:1981iu}. Since our results for the
beta functions of these theories differ from previous expressions in
the literature we feel it is worthwhile to present our results.

\section{Preliminaries}
\label{sec:theory}
The theory we study is given by\footnote{We have only written one chiral fermion, which would lead to an anomaly in the gauge symmetry. This is for simplicity; the potential anomaly will play no role in our work. In our computations below we will initially discuss the contribution of one fermion before generalizing to arbitrary matter content.}
\begin{equation}
\label{HDlag}
{\cal L}= -\frac12\Tr(F^{\mu\nu}F_{\mu\nu})
+\frac{1}{m^2}\Tr(D^\mu F_{\mu\nu})^2
+ \bar \psi_L i \slashed{D}\psi_L 
 +\frac{ \sigma_1}{m^2} \bar \psi_L 
 i\slashed{D}\slashed{D}\slashed{D} \psi_L 
 -\phi^* D^2 \phi -\frac{\delta_1}{m^2}\phi^*(D^2)^2 \phi  .
\end{equation}
Our notation is as follows. $A^a_\mu$ is the gauge field with field
strength $F_{\mu\nu}^a=\partial_\mu A^a_\nu+\cdots$. In matrix
notation $F_{\mu\nu}=T^a F_{\mu\nu}^a$ and $A_\mu=T^aA^a_\mu$ with
$T^a$ hermitian generators of the defining representation of the gauge
group (traceless for factors of a semisimple group). The `Tr' denotes
a trace in the space of these matrices,  the normalization is
$\Tr T^aT^b=\frac12\delta^{ab}$ and the structure constants are $[T^a,T^b]=if^{abc}T^c$. The covariant derivative is
$D_\mu=\partial_\mu\mathbf{1} +igA_\mu$. $\psi_L$ is a left handed
spinor, $\phi$ a complex scalar. Our metric convention is $(+---)$.

This is not the most general gauge and Lorentz invariant Lagrangian
with operators of dimension no larger than six, on two counts. First, we have omitted a scalar potential and Yukawa
scalar-spinor interactions. There is no technical barrier to
considering these, but our interest here is on the renormalization of
the gauge sector. And secondly, we extended the
renormalizable Lagrangian by three specific higher
derivative terms. These are the only terms one may add such that the
Lagrangian can be equivalently formulated as a renormalizable theory
that includes additional negative metric Lee-Wick (LW) fields. We
are particularly interested in this class of theories since it has
been shown that for massive vector scattering they preserve perturbative unitarity, while this is
not the case for theories with other type of dimension six derivative
operators\cite{Grinstein:2007iz}. 

Consider the Lagrangian,
\begin{multline}
\label{LWlag}
{\cal L}_{LW}= -\frac12\Tr(F^{\mu\nu}F_{\mu\nu})
+2\Tr( F_{\mu\nu} D^\mu\tilde A^\nu)-m^2\Tr \tilde A^\mu\tilde A_\mu\\
+ \overline \psi_L i \slashed{D}\psi_L +\overline {\tilde\psi}_L i
\slashed{D}\psi_L + \overline \psi_L i \slashed{D}{\tilde\psi}_L
- \overline {\tilde\psi}_R i \slashed{D}\tilde\psi_R  +\frac{m^2}{
  \sigma_1}\left( \overline {\tilde\psi}_L\tilde\psi_R+\overline
  {\tilde\psi}_R\tilde\psi_L\right) \\
-\phi^* D^2 \phi -\tilde\phi^* D^2 \phi-\phi^* D^2 \tilde\phi+\frac{m^2}{\delta_1}\tilde\phi^*\tilde\phi .
\end{multline}
Upon solving the equations of motion of the fields $\tilde A_\mu$, $\tilde\psi$ and
$\tilde\phi$ and inserting the solutions in \eqref{LWlag} one
recovers the higher derivative Lagrangian of \eqref{HDlag}. While the
Lagrangian \eqref{LWlag} has twice as many fields as the higher
derivative version \eqref{HDlag} it is renormalizable by power
counting, so it is more convenient to use in some cases. The mixing
terms present in~\eqref{LWlag} can be diagonalized by an appropriate
redefinition of the fields, as discussed in~\cite{us}.

For our calculations below we use the background field gauge (BFG)
method. Let us briefly review it. This is not only for completeness: as we
shall shortly show, one has to be careful about introducing higher
derivatives  in the gauge fixing term. Denote the quantum fields by
$A_\mu$ and the background fields by $B_\mu$. The effective action is
determined by the vacuum graphs for the theory with action integral
$S(A+B)$, where $S=\int d^4x {\cal L}$, and ${\cal L}$ as given
above. The gauge fixing condition is
\beq
{\cal F}(A,B)=0
\eeq
for some function that is invariant under gauge transformations of the
$B$ field with the $A$ field transforming as a matter field:
\begin{align}
\label{inv1}
B_\mu &\to U(\frac1{ig}\partial_\mu+B_\mu)U^\dagger\\
\label{inv2}
 A_\mu &\to UA_\mu U^\dagger .
\end{align}
The simplest covariant choice is
\beq
{\cal F}(A,B)=D(B)_\mu A^\mu=\partial_\mu A^\mu +ig[B_\mu,A^\mu] .
\eeq
By shifting $A\to A-B$ the BFG method can be understood in terms of the
formulation in the absence of a background field but with a
$B$-dependent gauge fixing condition,
\beq
{\cal F}(A-B,B)=0 .
\eeq
The Faddeev-Popov determinant $\Delta_{FP}$ can be computed through the ghost Lagrangian
\beq
{\cal L}_{FP}=\bar b[ D(B)^\mu D(A+B)_\mu] c ,
\eeq
where $b$ and $c$ are the anti-commuting scalars in the adjoint representation.

The gauge fixing condition can be brought into the action in the
functional integral by the usual trick: writing the condition as
$\delta({\cal F}-\alpha)$, one then uses the averaged partition
function
\beq
\label{exptrick}
Z=\int[d\alpha] \exp\left(\frac{-i}{2\xi}\int d^4x \alpha^2\right) Z_\alpha ,
\eeq
where 
\beq
Z_\alpha =\int[dA] e^{iS}\Delta_{FP}\delta({\cal F}-\alpha) .
\eeq
It is sometimes useful to have more derivatives in the gauge fixing
term in the action (for example, for power counting arguments). This
can be done by putting derivatives in the exponent in the
exponentiation trick \eqref{exptrick}. However we must be careful to
preserve the invariance in \eqref{inv1}--\eqref{inv2}. So an
alternative form of the partition function we may use is
\beq
\label{exptrick2}
Z=\sqrt{\det\left(1+\frac1{M^2}D(B)^2\right)} \int[d\alpha]
\exp\left(\frac{-i}{2\xi}\int \!d^4x\,
  \alpha\big[1+\frac1{M^2}D(B)^2\big]\alpha \right) Z_\alpha  .
\eeq
Notice the factor of the square root of the determinant,
which compensates for the extraneous $B$ dependence introduced
by the $\alpha$ integration.
Below we compute the beta functions
of this theory with both types of gauge fixing and find agreement.

The determinant in \eqref{exptrick2} can be computed using ghost fields,
 \beq
\det(1+\frac1{M^2}D(B)^2)=\int[db][dc]e^{i\int d^4x \bar
  b(-D(B)^2-M^2)c} .
\eeq 
The Lagrangian for these ghosts is similar to that of the Faddeev-Popov
ghosts, except that this one  has
a mass and lacks a coupling to the quantum field. This observation is
useful because as far as the computation of the infinite part of the
$B$ self-energy is concerned there is no difference between this  and the Faddeev-Popov
case. So these ghosts contribute to the infinite part of the
self-energy one half of the FP ghosts, the factor of  one half  arising from
taking  square root of the determinant.

One last comment is in order before we embark on our computation. To
properly construct the S-matrix for a Lee-Wick theory one needs to
adopt specific prescriptions for the choice of contours in Feynman
diagrams. As a result, there are well known difficulties in writing
a functional integral version of the quantization of the theory and
no consensus on whether a functional integral version exists; see
\cite{Boulware:1983vw,Kuti,van Tonder:2006ye}.  The above discussion on
the BFG method uses extensively the functional integral formalism. This
can be easily justified. To the extent that we are only interested in
renormalization, that is, in the ultraviolet divergences of the theory,
the detailed choice of integration contours is irrelevant. The difference
between any two integration contours in the complex energy plane in a
Lee-Wick amplitude gives always a residue at a pole, and is therefore
finite (even after integrating over spatial momentum).

\section{Renormalization}
The renormalized version of the Lagrangian \eqref{HDlag} is
\begin{multline}
\label{renLag}
{\cal L}= -\frac12Z\Tr(F^{\mu\nu}F_{\mu\nu}) +Z_\psi \bar \psi_L i \slashed{D}\psi_L-Z_\phi\phi^* D^2 \phi 
+\frac{1}{m^2}ZZ_{m^2}\left[\Tr(D^\mu
F_{\mu\nu})^2)\right] \\
 +\frac{1}{m^2}Z_\psi Z_{m^2}(Z_\sigma\sigma_1) \;\bar \psi_L \!\left[
 i \slashed{D}\slashed{D}\slashed{D}  \right]\!\psi_L 
 -\frac1{m^2}Z_\phi Z_{m^2}(Z_\delta\delta_1)\;  \phi^*\!\left[(D^2)^2 \right]\!\phi .
\end{multline}
The first line contains the kinetic terms (dimension four operators) and it
is in terms of these that the wave function renormalization factors $Z$,
$Z_\psi$ and $Z_\phi$ are defined. The next three lines contain the dimension
six operators for gauge fields, spinors and scalars, respectively. The coupling
constant renormalization is not shown explicitly, but it should be
understood that the Lagrangian depends on $g$ through the combination
$Z_gg$ only.

Some comments are in order. There are no counterterms of the form of
any of the dimension six operators, a result that was established in~\cite{us} by the power
counting analysis and verified through an explicit one loop
computation. This implies for example that $ZZ_{m^2}$ is finite, so we
can adopt the renormalization condition
\beq
\label{Zm}
ZZ_{m^2}=1 .
\eeq
Similarly, we have
\beq
\label{WF2}
Z_\psi Z_{m^2}Z_\sigma =Z_\phi Z_{m^2}Z_\delta=1 .
\eeq

We have chosen to work in background field gauge.  One of the great
simplifications of BFG is that~\cite{Abbott:1980hw}
\beq
\label{Zg}
Z_gZ^{\frac12}=1 .
\eeq
With this and Eqs.~\eqref{Zm}--\eqref{WF2} we deduce that the full set of
renormalization constants is given in terms of the three wavefunction
renormalization constants.

To write the Renormalization Group Equations (RGE) we Taylor expand the
renormalization constants with respect to $\epsilon\equiv 4-D$. Define residues
$a$ through
\begin{align}
Z &= 1+ \frac{a}{\epsilon}+\cdots\\
Z_g &= 1+ \frac{a_g}{\epsilon}+\cdots\\
&~ ~  ~\vdots\\
Z_\delta  &= 1+ \frac{a_\delta }{\epsilon}+\cdots .
\end{align}
Then, as usual,
\beq
\beta(g,\epsilon)=-\frac12\,\epsilon g +\beta(g), \qquad \beta(g)=\frac12\,g^2\frac{\partial
  a_g}{\partial g} .
\eeq
Using, from \eqref{Zg}, $a_g=-\frac12a$ and putting
together the contributions to the YM self energy in the previous
section we have,
\beq
\beta(g)=-\frac14\,g^2\frac{\partial a}{\partial g} .
\eeq
The anomalous dimensions for the matter fields are 
\beq
\gamma_f(g)=\frac12\mu \frac{\partial}{\partial\mu} \log Z_f =-\frac14 g\frac{\partial a_f}{\partial g},\quad
f=\psi,\phi  .
\eeq
The renormalization group equation  for the matter couplings is easily
obtained. We present this for a single spinor or scalar, to avoid
unnecessary complications from the matrix
structure:
\begin{align}
\mu\frac{\partial\sigma_1}{\partial\mu} &=-\sigma_1\gamma_{\sigma_1}(g)
=2\sigma_1\left(\gamma_\psi(g)-\frac{\beta(g)}{g}\right)\\
\mu\frac{\partial\delta_1}{\partial\mu}
&=-\delta_1\gamma_{{}_{\delta 1}}(g)
=2\delta_1\left(\gamma_\phi(g)-\frac{\beta(g)}{g}\right) ,
\end{align}
or more simply
\beq
\label{eq:running}
\mu\frac{\partial(g^2\sigma_1)}{\partial\mu}
=2(g^2\sigma_1) \gamma_\psi(g)
\qquad\text{and}\qquad
\mu\frac{\partial(g^2\delta_1)}{\partial\mu}
=2(g^2\delta_1) \gamma_\phi(g).
\eeq
We turn now to the explicit computation of the self-energy diagrams.

\subsection{YM self-energy}
\begin{figure}[htbp] 
   \centering
   \includegraphics[width=2in]{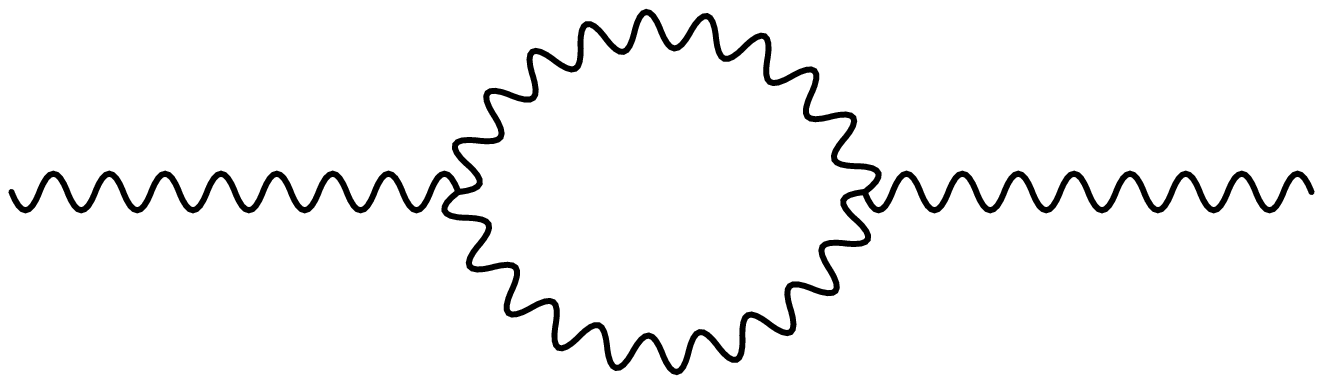} 
\hspace{1cm}   \includegraphics[width=2in]{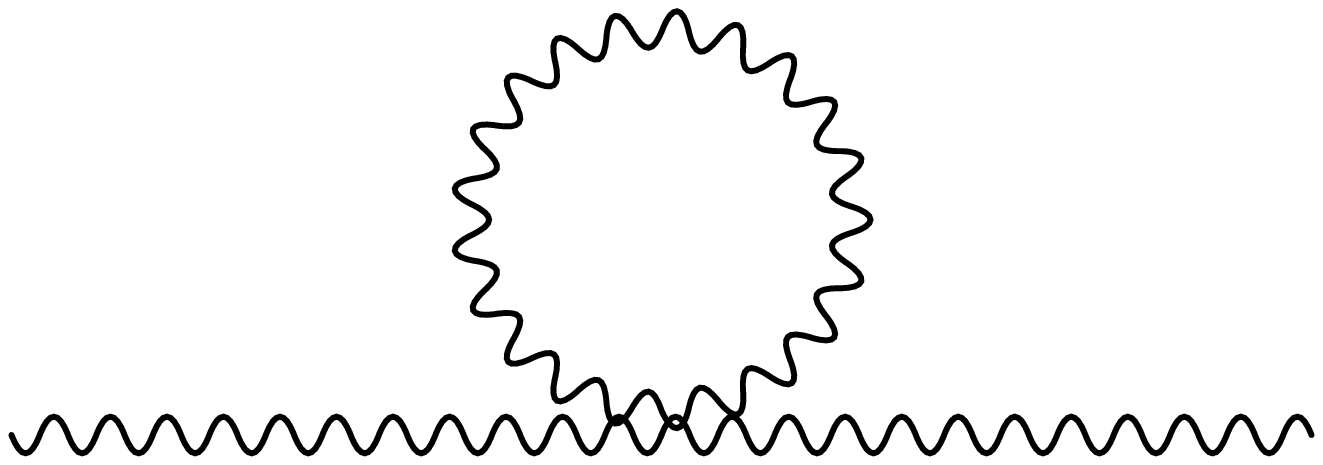} 
   \caption{Contribution to the self-energy of YM fields from internal YM fields}
   \label{fig:YMloop1}
\end{figure}

\begin{figure}[htbp] 
   \centering
   \includegraphics[width=2in]{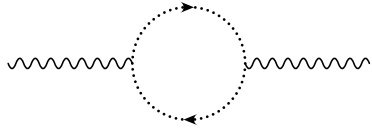} 
   \caption{Contribution to the self-energy of YM fields from internal ghosts}
   \label{fig:YMghost1}
\end{figure}

As mentioned above, we have performed the computation several
different ways: we can use a higher derivative version of the theory
with a standard covariant gauge fixing term, or we can use a higher
derivative version of the covariant gauge fixing term with a
Jacobian correction, or we can use the formulation of the theory
without higher derivative terms but instead including negative norm LW
fields. In each case the computation is very different. There is no
one to one correspondence between the contributions to the
renormalization constants of individual Feynman diagrams, yet the
resulting beta functions are the same. 

We first list our results for the $\epsilon$-poles of the graphs computed
in the higher derivative theory with a standard covariant BFG-term. The
graphs in Figs.\ref{fig:YMloop1} give
\begin{equation}
\label{YMloop1}
\frac{ig^2}{16\pi^2}\delta^{ab}\left(\frac2{\epsilon}\right)
\frac{41}6 C_2 (g_{\mu\nu}k^2-k_\mu k_\nu) ,
\end{equation}
where $C_2$ is defined by $\sum_{x,y}f^{axy}f^{bxy}=C_2\delta^{ab}$. 
The ghost graph in Fig.~\ref{fig:YMghost1} yields a contribution
\begin{equation}
\label{1loop-ghost}
\frac{ig^2}{16\pi^2}\delta^{ab}\left(\frac2{\epsilon}\right) C_2 \left(\frac13\right)(g_{\mu\nu}k^2-k_\mu k_\nu) .
\end{equation}

\begin{figure}[htbp] 
   \centering
   \includegraphics[width=2in]{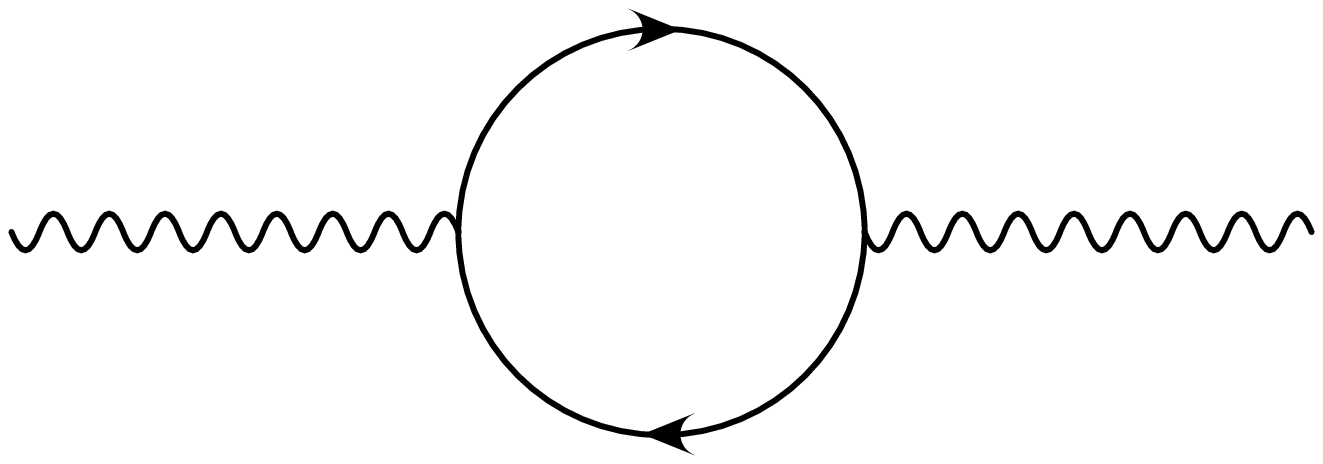} 
  \hspace{1cm} \includegraphics[width=2in]{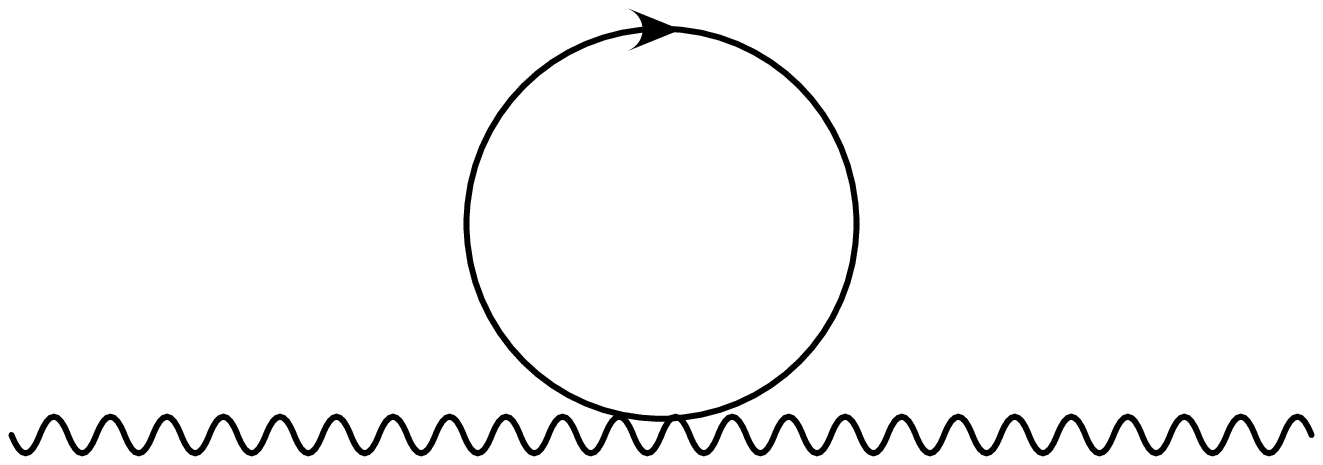} 
   \caption{Contribution to the self-energy of YM fields from internal spinor fields}
   \label{fig:YM-spinor1}
\end{figure}
\begin{figure}[htbp] 
   \centering
   \includegraphics[width=2in]{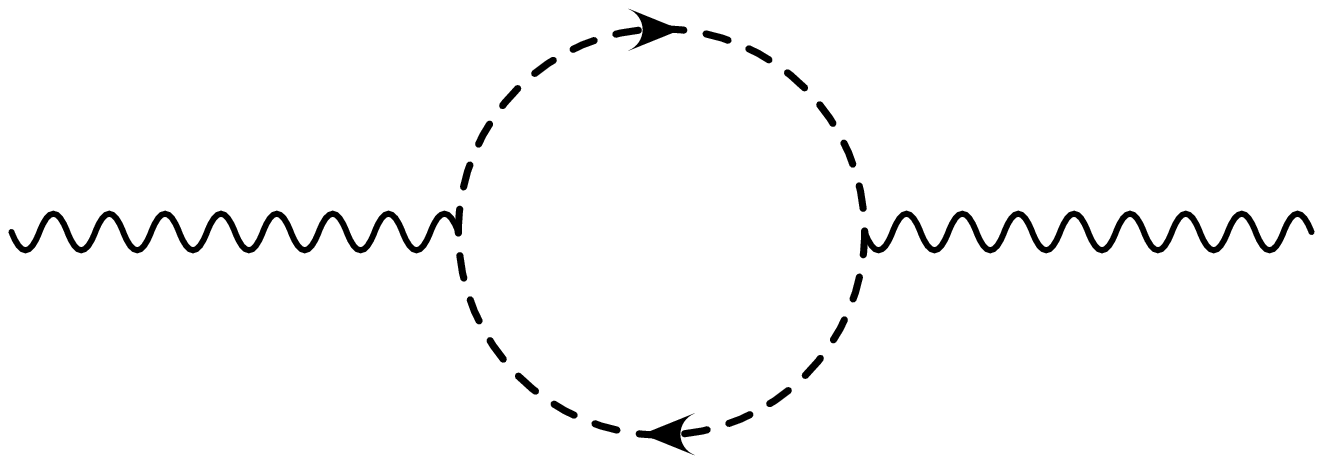} 
\hspace{1cm}   \includegraphics[width=2in]{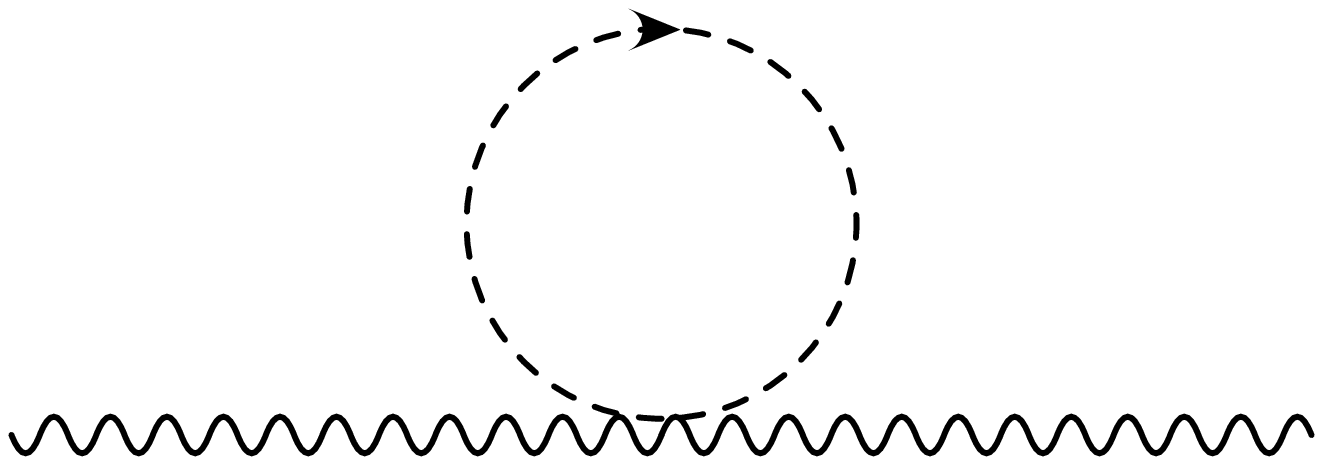} 
  \caption{Contribution to the self-energy of YM fields from internal complex scalar fields}
   \label{fig:YM-scalar1}
\end{figure}
Next come the matter fields. The spin-1/2 contribution (in the
fundamental representation of the gauge group) from
Figs.~\ref{fig:YM-spinor1} is 
\begin{equation}
\label{1loop-spinor}
-\frac{ig^2}{16\pi^2}\delta^{ab}\left(\frac2{\epsilon}\right) 
(g_{\mu\nu}k^2-k_\mu k_\nu) .
\end{equation}
Finally, the contribution from a complex scalar field (in the fundamental
representation) in Figs.~\ref{fig:YM-scalar1} is given by
\begin{equation}
\label{1loop-scalar}
-\frac{ig^2}{16\pi^2}\delta^{ab}\left(\frac2{\epsilon}\right) 
\frac13(g_{\mu\nu}k^2-k_\mu k_\nu).
\end{equation}

Now we turn to the case where we use a higher derivative version of the
covariant BFG-term. The only difference from the above is in the graphs
in Figs.\ref{fig:YMloop1} which now give
\begin{equation}
\frac{ig^2}{16\pi^2}\delta^{ab}\left(\frac2{\epsilon}\right)
\frac{40}6 C_2(g_{\mu\nu}k^2-k_\mu k_\nu) .
\end{equation}
However, now must also include a factor to compensate for
the background field dependence of the modified measure, see
Eq.~\eqref{exptrick2}. The determinant can be computed using a ghost as
explained in Sec.~\ref{sec:theory}, and the result is therefore 1/2 of
the usual ghost contribution of \eqref{1loop-ghost}, namely
\begin{equation}
\frac{ig^2}{16\pi^2}\delta^{ab}\left(\frac2{\epsilon}\right) \frac16 C_2 (g_{\mu\nu}k^2-k_\mu k_\nu) .
\end{equation}
The sum of these two contributions precisely equals the result in
\eqref{YMloop1}. 

Finally, we have computed the beta function in the Lee-Wick formulation of
the theory, as discussed in~\cite{us}. In this formulation, the physical
degrees of freedom are the gauge fields $A^a_\mu$, massive LW vector
fields $\tilde A^a_\mu$, a chiral spin $1/2$ field, a Dirac Lee-Wick
fermion, a scalar field and a Lee-Wick scalar field.  We
compute the beta function by computing the wavefunction renormalization
of the normal gauge fields in background field gauge.

\begin{figure}[htbp] 
   \centering
   \includegraphics[width=2in]{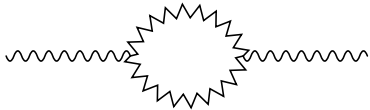} 
  \caption{Contribution to the self-energy of YM fields from internal
    LW-vector  fields}
   \label{fig:YMloop3}
\end{figure}

It is easy to deduce the contributions of the matter fields to the
beta function, because the LW fields couple to the gauge fields just
as normal fields do\footnote{Signs associated with Lee-Wick propagators
appear squared in all the relevant diagrams.}. Thus, the total contribution of the spin-$1/2$ fields
to the beta function is three times the usual contribution of a fundamental
chiral spin-$1/2$ fermion, while the scalar fields contribute twice the
usual scalar field value, in agreement with Eq.~\eqref{1loop-spinor}
and Eq.~\eqref{1loop-scalar}, respectively. It remains to compute the
effects of the LW vector fields. The relevant graph is shown in Fig.~\ref{fig:YMloop3}.  The graph evaluates to\footnote{In this formulation of the theory,
there are additional divergences proportional to $p^4$ and $p^6$ which
we do not show. These higher divergences are gauge artifacts. Since the
beta function is gauge independent to this order, we can be confident of
our results. It is possible to fix the gauge in the Lee-Wick formulation
of the theory so that these spurious divergences do not appear, at the
expense of a more involved formalism.}
\begin{equation}
\label{1loop:LWvector}
\frac{ig^2}{16\pi^2}\delta^{ab}\left(\frac2{\epsilon}\right)
\frac 7 2 C_2 (g_{\mu\nu}k^2-k_\mu k_\nu) .
\end{equation}
Of course, the gauge fields and ghost lead to a term
\begin{equation}
\frac{ig^2}{16\pi^2}\delta^{ab}\left(\frac2{\epsilon}\right)
\frac{11} 3 C_2 (g_{\mu\nu}k^2-k_\mu k_\nu) .
\end{equation}
in the beta function; adding this value to \eqref{1loop:LWvector} again
equals the result of \eqref{YMloop1}.

\subsection{Matter self-energies}
\begin{figure}[htbp] 
   \centering
   \includegraphics[width=2in]{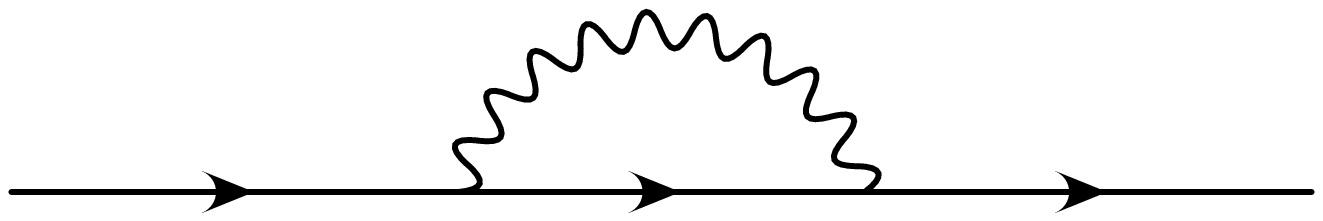} 
\hspace{1cm}   \includegraphics[width=2in]{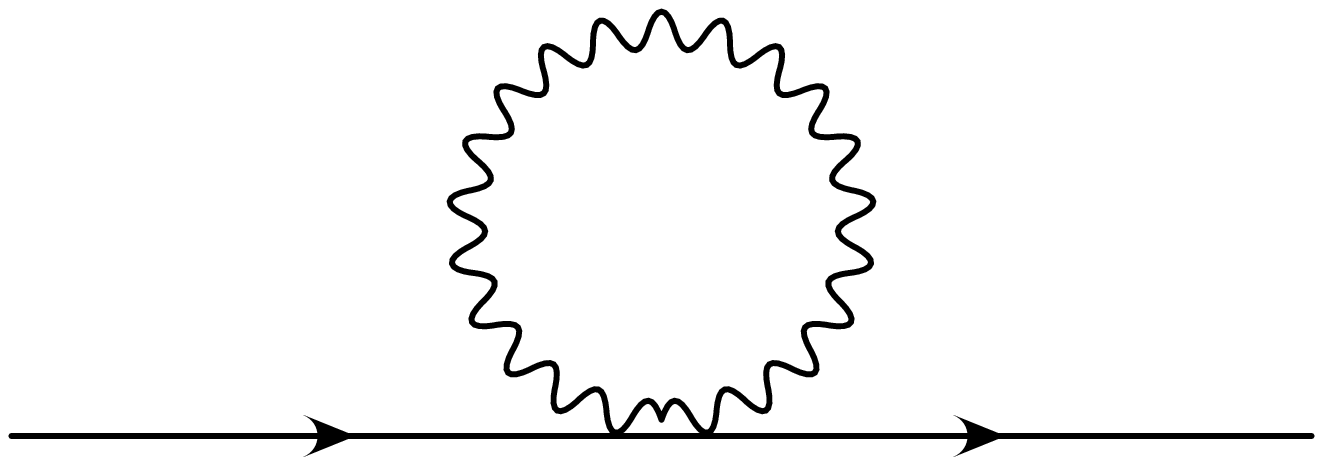} 
   \caption{Feynman diagrams contributing to the self-energy of spin-1/2 fields}
   \label{fig:spinor1}
\end{figure}

Now we turn to the self-energies of matter fields. The spin-1/2 self
energy is from Fig.~\ref{fig:spinor1}. We find the divergent terms
cancel among the two graphs. This result is the same for the higher
derivative theory, with either type of gauge fixing, as for the LW fields
version of the theory. 
\begin{figure}[htbp] 
   \centering
   \includegraphics[width=2in]{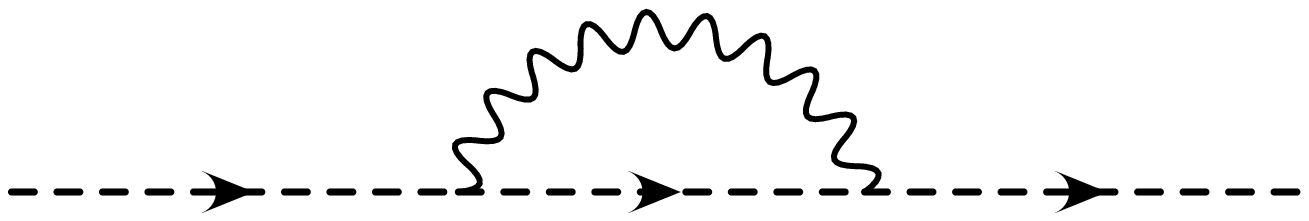} 
\hspace{1cm}   \includegraphics[width=2in]{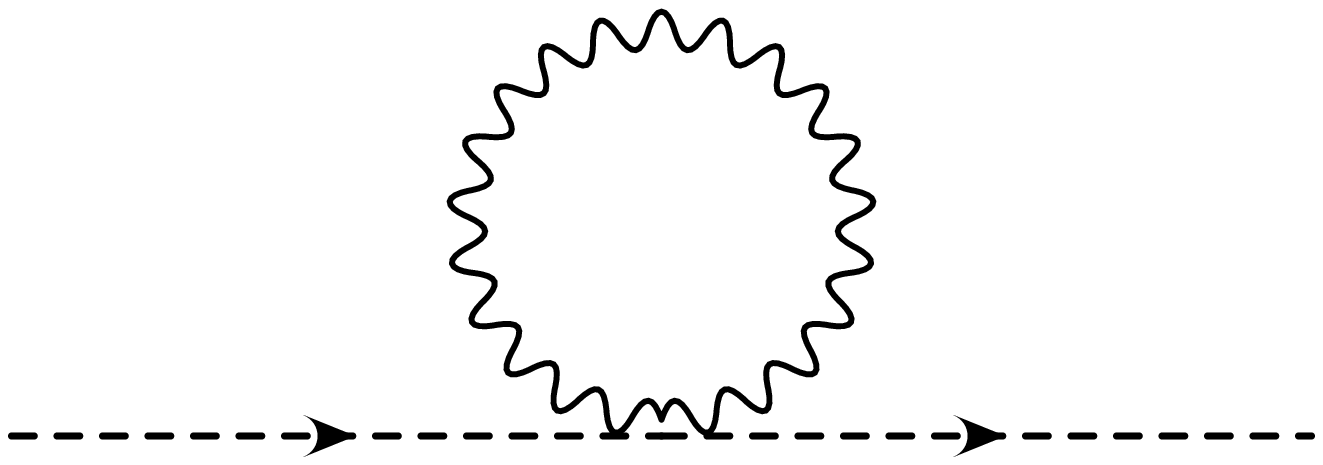} 
   \caption{Feynman diagrams contributing to the self-energy of spin-0 fields}
   \label{fig:scalar1}
\end{figure}
The  self-energy of the complex scalar from the sum of the diagrams in
Fig.~\ref{fig:scalar1} gives a wavefunction renormalization 
\begin{equation}
\frac{g^2}{16\pi^2}\left(\frac2{\epsilon}\right) 6\,C_1 \delta_1
ik^2 
\end{equation}
where $C_1$ is defined by $T^aT^a=C_1\mathbf{1}$. There is of course
also mass renormalization but recall we have postponed the study of the
renormalization of terms in the scalar potential. 

\section{Beta Function and Anomalous Dimensions}
Our final results are in the form of explicit expressions for the beta
functions and anomalous dimensions. These are obtained combining the
results above. First, the running of the gauge coupling is determined by
\beq
\beta(g)=-\frac{g^3}{16\pi^2}\left[\frac{43}{6}C_2
  -n_f-\frac13n_s\right] .
\eeq
We have introduced $n_f$ and $n_s$ for the number of spinor and scalar
fields. More generally, if
the spin 1/2 and 0 fields are in arbitrary representations of the
gauge group we have
\beq
\label{beta-LW}
\beta(g)=-\frac{g^3}{16\pi^2}\left[\frac{43}{6}C_2
  -2\sum_f n_fT(f)-\frac23\sum_sn_sT(s)\right] ,
\eeq
where in the representation $x$ we have $\Tr(T^a T^b)=T(x)\delta^{ab}$.

Similarly, the anomalous dimensions for the spinor and scalar are
\beq
\gamma_\psi(g)=0 ,
\eeq
and
\beq
\gamma_\phi(g)=-\frac{g^2}{16\pi^2} 3C_1\delta_1 .
\eeq

Combining \eqref{Zm} and \eqref{Zg} we see that $Z_{m^2}=Z_g^2$, so
the solution to the RGE for $m^2$ is immediate,
\beq
m^2(\mu)=\left(\frac{g^2(\mu_0)}{g^2(\mu)}\right)m^2(\mu_0) .
\eeq
Since $\gamma_\psi(g) = 0$, we see from Eq.~\eqref{eq:running} that
$g^2 \sigma_1$, and so $m^2 / \sigma_1$ does not run. Therefore, the
mass of the Lee-Wick fermion is an invariant of the RG flow. On the other
hand, the quantity $g^2 \delta_1$ does flow so that the LW scalar mass
grows logarithmically in the ultraviolet.

The result \eqref{beta-LW} is roughly what one would guess
naively. The higher derivative terms that we have introduced in the
Lagrangian are precisely the ones that can be described as additional
LW fields. Hence one roughly expects to double the contribution
of each field to the $\beta$-function. Subtleties occur in the spinor
matter and pure gauge terms. In the spinor terms, the contribution is
tripled because the Lee-Wick partner of a chiral fermion is
non-chiral. In the pure gauge term, the contribution from ghosts is not
quite doubled as explained above.

Hence, much like for the standard model of electroweak interactions, the
LW extension of the standard model does not display good unification of
coupling constants. The standard model does however unify well if properly
chosen additional fields are introduced. A simple example was given by
Willenbrock in Ref.~\cite{Willenbrock:2003ca}, where he shows that the
standard model with six Higgs doublets unifies. Similarly, we find that
the Lee-Wick extension of the standard model has good coupling constant
unification if it is extended to include six or seven Higgs doublets.

However, Willenbrock points out that in the six-Higgs doublet model the
unification scale is very low so if the unification group is simple,
then the proton decays excessively fast. He proposes an interesting
solution to this problem using trinification, that is, a unified group
$SU(3)^3/Z_3$. The unification scale in our six Higgs doublet model
is even lower than in Willenbrock's case, about a million times the LW
scale $m$. Presumably one can formulate a LW extension of trinification,
but we have not pursued this.

\section{Additional Dimension Six Operators}

In this section we consider a more general theory which contains
additional dimension six operators in the Lagrangian density. This 
theory does not satisfy the constraints of
perturbative unitarity, so that scattering amplitudes  cannot
be computed by perturbative methods. The beta function and anomalous
dimensions, on the other hand, may still be computed in
perturbation theory since no large energies occur in these
functions. Theories of these types have been considered in the
literature previously as toy models for higher derivative gravity~\cite{Fradkin:1981hx,Fradkin:1981iu}.
The Lagrangian of the theory is given by
\begin{equation}
\mathcal{L} = \mathcal{L}_A + \mathcal{L}_\psi + \mathcal{L}_\phi , 
\end{equation}
where
\beq
{\cal L}_A = -\frac12\Tr(F^{\mu\nu}F_{\mu\nu})
+\frac{1}{m^2}\Tr(D^\mu F_{\mu\nu})^2-\frac{i\gamma g}{m^2}\Tr(F^{\mu\nu}[F_{\mu\lambda},F_\nu^{\phantom{\nu}\lambda}]) 
\eeq
specifies the dynamics of the gauge sector. The spinor matter Lagrangian is
\beq
{\cal L}_\psi = \bar \psi_L i \slashed{D}\psi_L +\frac{i}{m^2} \bar \psi_L \left[
  \sigma_1\slashed{D}\slashed{D}\slashed{D}+\sigma_2\slashed{D}D^2+ig\sigma_3F^{\mu\nu}\gamma_\nu D_\mu 
+ig\sigma_4 (D_\mu F^{\mu\nu})\gamma_\nu  \right]\psi_L ,
\eeq
where, in the last term of the Lagrangian, the covariant derivative acts only on the field
strength tensor, and $\sigma_{1-4}$ are dimensionless constants. 
For a complex scalar, we consider for the Lagrangian density
\beq
{\cal L}_\phi = -\phi^* D^2 \phi -\frac1{m^2}\phi^*\left[\delta_1(D^2)^2 +ig
  \delta_2(D_\mu F^{\mu\nu})D_\nu +g^2 \delta_3 F^{\mu\nu}F_{\mu\nu}\right]\phi ,
\eeq
where, as above, the parenthesis in the second term indicates that the
derivative to the left of $F_{\mu\nu}$ acts only on  $F_{\mu\nu}$. 

We find that the beta function and anomalous dimensions are given by
\begin{align}
\label{beta-extended}
\beta(g)&=-\frac{g^3}{16\pi^2}\left[\left(\frac{43}{6}-18\gamma+\frac92\gamma^2\right)C_2
  -n_\psi \left(\frac{\sigma_1^2-\sigma_2\sigma_3+\frac12\sigma_3^2}{(\sigma_1+\sigma_2)^2}\right)
-n_\phi \left(\frac{\delta_1+6\delta_3}{3\delta_1}\right)\right],  \\
%
\gamma_\psi(g)&=-\frac{g^2}{16\pi^2} \frac34C_1
\left(\frac{2\sigma_1(2\sigma_2+\sigma_3-2\sigma_4)+\sigma_2(2\sigma_2+2\sigma_3-\sigma_4)-\sigma_3^2-\sigma_4^2+\sigma_3\sigma_4}{\sigma_1+\sigma_2}\right) , \\
\gamma_\phi(g)&=-\frac{g^2}{16\pi^2} \frac38C_1
\left(\frac{8\delta_1^2-\delta_2^2-4\delta_1\delta_2}{\delta_1}\right) .
\end{align}
We note that our expression for the beta function differs from that found in Appendix C
of~\cite{Fradkin:1981iu}. We can write the beta functions for the couplings of the 
dimension six operators in terms of these anomalous dimensions. The first states
that $\gamma$ is a constant,  
\beq
\mu\frac{\partial\gamma}{\partial\mu}=0.
\eeq
The equations for $\sigma_i$ and $\delta_i$ are  the same as we found
for $\sigma_1$ and $\delta_1$ in the previous section, 
\beq
\mu\frac{\partial(g^2\sigma_i)}{\partial\mu}
=2(g^2\sigma_i) \gamma_\psi(g)
\qquad\text{and}\qquad
\mu\frac{\partial(g^2\delta_i)}{\partial\mu}
=2(g^2\delta_i) \gamma_\phi(g).
\eeq
In particular, we see that the ratios $\sigma_i/\sigma_j$ and
$\delta_i/\delta_j$ do not run.

Clearly the renormalization group of this theory is much richer than
the one considered in the previous section. In particular, for the
theory based on the standard model (that is, the theory which has the same field
content as the standard model), there is now an additional free parameter that enters the
scale of unification, namely the cubic field strength coupling of the
unified theory, $\gamma$. One can in fact have successful $SU(5)$
unification in this theory, with a unification scale in excess of
$10^{16}m$ for $0.33 \lesssim \gamma \lesssim 0.35$ or $ 3.65\lesssim
\gamma \lesssim 3.67$. While this may seem phenomenologically
adequate, we remind the reader that this theory is not perturbatively
unitary.

Another interesting property of the result in \eqref{beta-extended} is
that the coupling constants, $\gamma$, $\sigma_i$ and $ \delta_i$ can
be chosen to make the $\beta$-function vanish. 
Note that the quantity in square brackets in
\eqref{beta-extended} is renormalization group invariant so one may
consistently set it to any fixed value.  

\section{Concluding Remarks}

Typically, Lagrangian densities in particle physics which contain
operators of dimension higher than four lead to theories which are less
predictive. This is a result of the divergences introduced by these
operators in perturbation theory. New counterterms must be introduced
to absorb these divergences, typically leading to theories containing
an infinite number of couplings constants, which are a priori unknown.

The situation is different in Lee-Wick theories. In the higher derivative
formulation of the theories, dimension six operators are present in the
microscopic Lagrangian. There is one new constant associated with each
higher derivative operator, which physically corresponds to the mass of
the corresponding Lee-Wick degree of freedom. 

In this work, we have described the renormalization of Lee-Wick theories
to one-loop order. No new counterterms are required to absorb the
divergences of the theory. In fact, we have shown that the wavefunction
renormalizations of the various fields present in Lee-Wick gauge theory
contain all the information about the renormalization group running of the
theory. For the Lee-Wick gauge bosons, we have shown that the quantity
$m^2 g^2$ is an invariant of the renormalization flow. Thus, the new
constant introduced in the definition of a Lee-Wick gauge theory truly
is just one number, and not a new function of energy scale. In addition,
we learn that if the theory is asymptotically free, then the LW vector
boson mass flows to infinity in the UV. This counter intuitive behaviour
is interesting because it indicates that the ultraviolet fixed point
of the RG flow of an asymptotically free pure Lee-Wick gauge theory is
a normal quantum field theory: the scale suppressing the dimension six
operator in the higher dimension formulation of the theory has become
infinite so that this term no longer contributes to the dynamics. The
remaining degrees of freedom are the usual gauge bosons.

We have obtained expressions for the beta function and anomalous
dimensions of scalar and spinor matter. The coupling runs more quickly in
Lee-Wick theory compared to the usual non-Abelian gauge theory. We find
that the Lee-Wick standard model does not unify naturally, and that,
on account of the more rapid running of the coupling, the unification
scale of the theory augmented with extra field content is typically
rather low.  In addition, we find that the anomalous dimension of spinor
matter vanishes.

Finally, we have discussed some more general theories containing dimension
six operators which are not of Lee-Wick type. Since amplitudes in these
theories grow too quickly with energy to satisfy perturbative unitarity
bounds, the theories either become non-perturbative at some scale,
or else they violate unitarity. However, no large factors of energy
appear in the expressions for the beta function or for the anomalous
dimensions, so they may still be computed in perturbation theory. (Of
course, they no longer give us insight into the high energy behaviour
of physical scattering amplitudes.) These theories have been discussed
elsewhere in the literature, and since our results differ from previous
expressions we have reported our results above. Our results
indicate that if it is possible to make sense of these theories, then,
for suitable choices of the couplings, these theories may enjoy the
property that their beta function vanishes.

\begin{acknowledgments}
We thank Mark B. Wise for useful discussions and collaboration at
the beginning of this work, and we are grateful to Arkady A. Tseytlin
for email correspondence. DOC thanks Poul Henrik Damgaard for several
helpful conversations. The work of BG and DOC was supported in part
by the US Department of Energy under contracts DE-FG03-97ER40546 and
DE-FG02-90ER40542, respectively.
\end{acknowledgments}

\end{document}